# Towards Quantum Communication in Free-Space Seawater

Ling Ji[1,2,†], Jun Gao[1,2,†], Ai-Lin Yang[1,2], Zhen Feng[1,2], Xiao-Feng Lin[1,2], Hong-Gen Li[1], and Xian-Min Jin[1,2,*]

[1]State Key Laboratory of Advanced Optical Communication Systems and Networks, Institute of Natural Sciences & Department of Physics and Astronomy, Shanghai Jiao Tong University, Shanghai 200240, China

[2]Synergetic Innovation Center of Quantum Information and Quantum Physics, University of Science and Technology of China, Hefei, Anhui 230026, China

[†]These authors contributed equally to this work.

**Long-distance quantum channels capable of transferring quantum states faithfully for unconditionally secure quantum communication[1,2,3] have been so far confirmed feasible in both fiber[4,5,6] and free-space air[7,8,9]. However, it remains unclear whether seawater, which covers more than 70% of the earth, can also be utilized, leaving global quantum communication incomplete. Here we experimentally demonstrate that polarization quantum states including general qubits and entangled states can well survive after travelling through seawater. We performed experiments in a 3.3-meter-long tube filled with seawater samples collected in a range of 36 kilometers in Yellow sea, which conforms to Jerlov water type I[10]. For single photons at 405 nm in blue-green window, we obtained average process fidelity above 98%. For entangled photons at 810 nm, even with high loss, we observe violation of Bell**

**inequality[11] with 33 standard deviations. This work confirms feasibility of seawater quantum channel, representing the first step towards underwater quantum communication.**

Underwater communication is vital for undersea exploitation and modern communication. Conventional ways which employ acoustical technique for underwater communication have their drawbacks[12,13], including high path loss, narrow bandwidth, high bit error rate, among which unconditional security is more demanding due to commercial and secure interest. The question arises of whether quantum communication[14,15] can be achieved by employing seawater as a reliable channel.

Since the seminal work of Bennett and Brassard on transferring quantum states and cryptographic keys through 0.3-meter-long free-space air[16], quantum communication has become an ultimate approach for the goal of unconditional communication security, with massive experimental efforts pushing the distance up to the order of 100 kilometers[17,18,19,20,21]. Although quantum key distribution and quantum teleportation have been achieved via optical fiber installed underneath Geneva Lake[4] and the River Danube[6] respectively, experimental investigation in free-space seawater has never been done so far.

Fortunately, as existence of transmission window around 800 nm in free-space air[9], there is a "blue-green" optical window at the wavelength regime of 400-500 nm in free-space seawater[22], wherein photons experience less loss and therefore can penetrate deeper. Photonic polarization may be a desirable carrier of quantum bits since isotropic seawater induces very limited birefringent effect. In this letter, to verify feasibility of seawater quantum channel, we experimentally explore polarization preservation properties of single photon and quantum entanglement.

As shown in Fig.1, Our seawater samples were collected from the surface of costal sea in the zone between Dalian city and Zhangzi island. Six sampling sites are several kilometers apart from one another. The distance between site *I* and *VI* is up to 36 kilometers. Investigations in such a large scale would ensure generalizability of our experimental results. A schematic layout of the experimental setup is shown in Fig. 2, a 3.3-meter-long glass tube equipped with inlet and outlet is utilized as a testbed of quantum channel by filling different seawater sample as well as distilled water. A 405-nm semiconductor laser running at CW mode is fed into the channel and detected by power meter. Together with a reference detection for eliminating laser fluctuation, we obtain average attenuation coefficient $0.354 \pm 0.007$ m$^{-1}$ in six seawater samples. The value is a few times higher than the result of $0.081 \pm 0.009$ m$^{-1}$ obtained in distilled water [see Methods] (Table 1).

Besides the loss associated with maximum secure distance, it is more important to find a desirable degree of freedom of photon with which quantum states can be encoded and transferred with high fidelity, i.e. without significant degrade induced by seawater. As water is a uniform isotropic medium, it is very likely that seawater, though incorporating dissolved salts and microbes, does not lead to massive polarization rotation or depolarization of photon. Multi random scattering on suspended particulate matter can introduce depolarization accompanying with transverse angle diffusion[23]. However, we can spatially filter the depolarized photons with small receiving angle defined by optical fiber.

We prepare single photons by driving the semiconductor laser to 2-ns pulse with a repetition rate of 50 MHz and strongly attenuating it to 0.3~0.6 photon per pulse typically adopted in decoy state protocols[24,25,26]. We encode single photons with six initial input states $|\psi_{in}\rangle$ in polarization $|H\rangle$, $|V\rangle$, $|D\rangle = \frac{1}{\sqrt{2}}(|H\rangle + |V\rangle)$, $|A\rangle = \frac{1}{\sqrt{2}}(|H\rangle - |V\rangle)$, $|R\rangle = \frac{1}{\sqrt{2}}(|H\rangle + i|V\rangle)$ and $|L\rangle = \frac{1}{\sqrt{2}}(|H\rangle - i|V\rangle)$ respectively, where $H$ and $V$ represent horizontal and vertical polarization. After going through seawater, we project each output state on these three sets of orthogonal basis (Fig. 2a). We employ quantum state tomography[27] method to reconstruct the density matrixes of output states with the counts of avalanche photodiode detectors APD1 and APD2. By making coincidence with synchronization signal of laser pulse from photodiode, we suppress noisy counts including background count and dark count of avalanche photodiode detectors from 500 cps to 150 cps, when the coincidence window of FPGA is 3.5 ns. Fig. 2b shows fidelities of receiving states for six initial states in six seawater samples, distilled water and air (empty tube), see their average value in Table 1. We can see that fidelities are all above 98%, which far exceed the classical limit 2/3[28]. As an example, Fig. 2c shows density matrix of six receiving states through seawater sample *VI*.

To reveal physical processes of seawater quantum channel, we employ quantum process tomography[29], which can provide complete information of the channel changing arbitrary input state into another state. Any quantum process can be written as

$$\rho_{out} = \sum_{m,n} \chi_{mn} A_m \rho_{in} A_n . \qquad (1)$$

Once the set of operators $A_m$ are fixed, the density matrix of process $\chi_{mn}$ can detail the dynamics of the system. We measure $\chi_{mn}$ with each sample and air. We derive the process fidelities as $F_P = \left\{ tr[\sqrt{\sqrt{\chi_{seawater}} \chi_{ideal} \sqrt{\chi_{seawater}}}] \right\}^2$. In spite of attenuation difference, we obtain high process fidelities in all seawater samples as well as distilled water (see Table 1).

Quantum entanglement is the main resource for quantum communication. The randomness and correlations inherent in quantum entanglement can be exploited to enable entanglement-based quantum cryptography[2], quantum teleportation[3], quantum repeater[30] and distributed quantum computing[31]. It is therefore of practical interest to see whether entanglement can be preserved in seawater channel. Furthermore, since the polarization state of entangled photons is completely uncertainty before detection, the survival of entanglement will reveal that seawater channel allow faithful transmission of arbitrary unknown polarization states, consistent with the result of high process fidelity of seawater channel.

In Fig. 3a, polarization entangled photon-pair source at 810 nm is generated by a blue laser beam (power 11mw, wavelength 405nm) pumping a quasi-phase matched periodically-poled KTiOPO$_4$ (PPKTP) crystal with type II spontaneous parametric down conversion[32]. The source is prepared as the singlet state

$$|\Psi^-\rangle = \frac{1}{\sqrt{2}}(|H_A\rangle|V_B\rangle - |V_A\rangle|H_B\rangle) \quad (2)$$

where the subscripts *A* and *B* label the spatial modes. Locally we obtain more than 300k cps in each path, and 55k cps coincidence events. Photon *B* is sent through the glass tube connected by a 3-meter-long single mode fiber. We analyze photon *A* locally and photon *B* at the receiving site with state tomography.

By filling seawater sample *VI* into the glass tube, we have observed a change of polarization correlation from 95.28±0.14% to 94.99±0.42% in *H/V* basis, from 94.15±0.16% to 94.6±0.44% in *D/A* basis (see Fig. 3b and 3c), indicating no clear decrease even under 30 dB additional loss at 810 nm wavelength. State tomography method is used to construct the density matrix of the entangled state under two conditions, i.e. air and sample *VI* (see Fig. 4). We have obtained a fidelity 0.9946 from the distance of these two density matrices by $F_S = \left\{ tr[\sqrt{\sqrt{\rho_{seawater}} \rho_{air} \sqrt{\rho_{seawater}}}] \right\}^2$.

We further evaluate the quantity of output entangled state via a clear violation of the Clauser-Horne-Shimony-Holt (CHSH) -type Bell's inequality[11], with regard to classical limit $S = 2$ and perfect entangled state $S = 2\sqrt{2} \sim 2.828$. We achieved $S = 2.6936 \pm 0.0074$ in air and $S = 2.6695 \pm 0.0203$ in seawater sample, violating CHSH-type Bell's inequality with 93 and 33 standard deviations respectively. Again, there is no distinctive difference in the two conditions, though confidence drops as the result of less coincidence events collected in seawater due to loss. These results indicate polarization entangled state can survive well in seawater channel.

Our results have verified free-space seawater desirable as quantum channel enabling high-fidelity distribution of single photon and entanglement. It encourages us to look into an achievable communication distance. The seawater farther from the coastline, which contains less suspended particulate matter and therefore has much lower attenuation, may promise longer transmission distance for photons. According to data reported[10,33], the loss to photons in blue-green window can be as low as 0.018 m$^{-1}$. An achievable communication distance 885 meters can be derived if we apply attainable threshold of quantum communication against loss of 70 dB[15]. In fact, there have been extensive applications once quantum communication with objects underwater 200 meters becomes feasible.

However, further experimental tests in the scenario of longer channel and field condition are necessary, where new techniques have to be developed to solve emerging problems. For example, we have observed slow beam wandering in our comparatively static ambient condition [see Methods]. It implies ATP, a technique of active locking of beam pointing via feedback, has to be developed according to real dynamic environment of open sea. Quantum repeater[30] combining entanglement swapping[34] and quantum memory[35] in blue-green window, though technically more challenging, can efficiently extend the achievable distances. As its elemental ingredients, short-wavelength quantum entanglement has been developed in semiconductor[36], and short-wavelength quantum memory may be achieved by virtue of frequency conversion in quantum regime[37,38,39].

In summary, we have experimentally demonstrated the distribution of polarization qubits and entangled photons over seawater channel. The high process fidelities indicate the seawater associated with suspended particulate matter introduces very limited depolarization, which verify the feasibility of quantum communication and quantum cryptography in free-space seawater. Future explorations include field experiment in open sea, blue-green band quantum repeater and air-sea quantum communication interface.

## Methods

**Loss budget and measurement.** Loss caused by seawater could not be measured directly because of additional loss of optical devices and possible fluctuation of the laser's power. We use a reference beam to remove the influence. Reference beam and signal beam are divided by a polarization beam splitter and for every sample the splitting ratio is constant. Real loss induced by seawater can be derived by subtract the value of empty tube out of full tube. Differential reflection at glass-seawater and glass-air interface must be taken into account, which can be figured out by Fresnel formula.

$$\tilde{t} = \frac{2n_1 \cos i_1}{n_1 \cos i_1 + n_2 \cos i_2}$$

$$T = \frac{n_2}{n_1} t^2$$

where $i_1$ ($i_2$) is incident angle (transmission angle). Refractive index of medium can influence the transmission of light. $n_{air} = 1$, $n_{water} = 1.34$ ($T_C = 20°C$, $\lambda = 405nm$), $n_{BK7} = 1.5302 (\lambda = 405nm)$. Seawater water is a complex system, whose refractive index is related to its salinity, temperature and the incident light wavelength[40], which can be written as

$$n_{seawater} = a_0 + (a_1 + a_2 T_C + a_3 T_C^2)S + a_4 T_C^2 + (a_5 + a_6 S + a_7 T)\lambda^{-1} + a_8 \lambda^{-2} + a_9 \lambda^{-3}$$

where empirical coefficients $a_1 \sim a_9$ are $a_0 = 1.31405$, $a_1 = 1.779 \times 10^{-4}$,

$a_2 = -1.05 \times 10^{-6}$, $a_3 = 1.6 \times 10^{-8}$, $a_4 = -2.02 \times 10^{-6}$, $a_5 = 15.868$, $a_6 = 0.01155$, $a_7 = -0.00423$, $a_8 = -4382$, $a_9 = 1.1455 \times 10^6$. Thus we can get the transmission rate $T_{air} = 83.57\%$ and $T_{seawater} = 90.535\%$ under the conditions that the tube is empty and full, respectively.

Assuming that the intensity of reference beam is $a_1(b_1)$ and the intensity of output signal beam is $a_2(b_2)$, corresponding to the condition empty tube (full tube), we can obtain attenuation coefficient by

$$\alpha = \log\left(b_2 \Big/ \left(\frac{a_2 b_1}{a_1} \times \frac{90.535\%}{83.57}\right)\right) \Big/ -3.3$$

The attenuation coefficient of Jerlov type I coastal water is $\alpha_{average} = 0.35667\,\mathrm{m}^{-1}$ at $\lambda = 425\,\mathrm{nm}$, which is close to the result we obtain in our experiment $\alpha_{average} = 0.35365 \pm 0.0068\,\mathrm{m}^{-1}$ at $\lambda = 405\,nm$.

**Differential efficiency in quantum tomography.** Quantum state tomography can experimentally determine density matrix of an unknown state with a set of projection measurements. Density matrix of unknown state of single photon can be represented by four parameters $\{S_0, S_1, S_2, S_3\}$ as $\hat{\rho} = \frac{1}{2}\sum_{i=0}^{3} S_i \sigma_i$. Due to normalization, $S_0 = 1$. The other parameters can be determined with projection measurements as $S_0 = P_{|H\rangle} + P_{|V\rangle}$, $S_1 = P_D - P_A$, $S_2 = P_{|R\rangle} - P_{|L\rangle}$, $S_3 = P_{|H\rangle} - P_{|V\rangle}$. For example, probe state $|H\rangle$ is sent through the seawater sample and output state is projected on $|H\rangle\langle H|$ ($|V\rangle\langle V|$), which is recorded by coincidence counts between photodiode and APD1 (APD2) denoted by $C_{H1}$ ($C_{H2}$). Stokes parameter $S_3$ can be determined by $C_{H1}$ ($C_{H2}$) and $C_{V1}$ ($C_{V2}$). The differential ratio of projection measurements $|H\rangle\langle H|$ and $|V\rangle\langle V|$ is $\eta_{HV} = \sqrt{\frac{C_{H1}C_{V1}}{C_{H2}C_{V2}}}$. To eliminate differential loss and detection efficiency, we correct the counts by

$S_3 = \frac{C_{H1} - C_{H2}\eta_{HV}}{C_{H1} + C_{H2}\eta_{HV}}$. By using the same method, we can get $\eta_{DA} = \sqrt{\frac{C_{D1}C_{A1}}{C_{D2}C_{A2}}}$,

$S_2 = \frac{C_{D1} - C_{D2}\eta_{DA}}{C_{D1} + C_{D2}\eta_{DA}}$ and $\eta_{RL} = \sqrt{\frac{C_{R1}C_{L1}}{C_{R2}C_{L2}}}$, $S_1 = \frac{C_{R1} - C_{R2}\eta_{DA}}{C_{R1} + C_{R2}\eta_{DA}}$.

**Beam wandering in seawater.** In the experiment, single count of detector present slow and quasi-periodical variation. We attribute this to the beam wandering associated with decoupling to optical fiber. Ambient condition change including mechanical vibration, temperature fluctuation and varying salinity may result in variational and inhomogenous refractive index distribution along propagating direction of the beam. According to the data collected in sample *VI*, the counts of two detectors present nearly synchronous variation. Such beam wandering issue, though in a faster speed, also exists as a main problem to solve in free-space quantum communication. It suggests that the active feedback system adopted in free-space air should also be applied to free-space seawater.

**Acknowledgements**

The authors thank J.-W. Pan, I. A. Walmsley, L.-J. Zhang and X.-F. Ma for helpful discussions. This work was supported by the National Natural Science Foundation of China under Grant No.11374211, the Innovation Program of Shanghai Municipal Education Commission (No.14ZZ020), Shanghai Science and Technology Development Funds (No.15QA1402200), and the open fund from HPCL (No.201511-01). X.-M.J. acknowledges support from the National Young 1000 Talents Plan.

**Author contributions**

X.-M.J. conceived the work and supervised the project. L.J., J.G., Z.F., A.-L.Y., X.-F.L., H.-G.L. and X.-M.J. all contributed to designing and setting up the experiment. L.J., A.-L.Y. and Z.F. collected seawater samples, L.J. and X.-M.J. analyzed the data and wrote the paper.

**Additional Information**

The authors declare that they have no competing financial interests. Correspondence and requests for materials should be addressed to the authors.

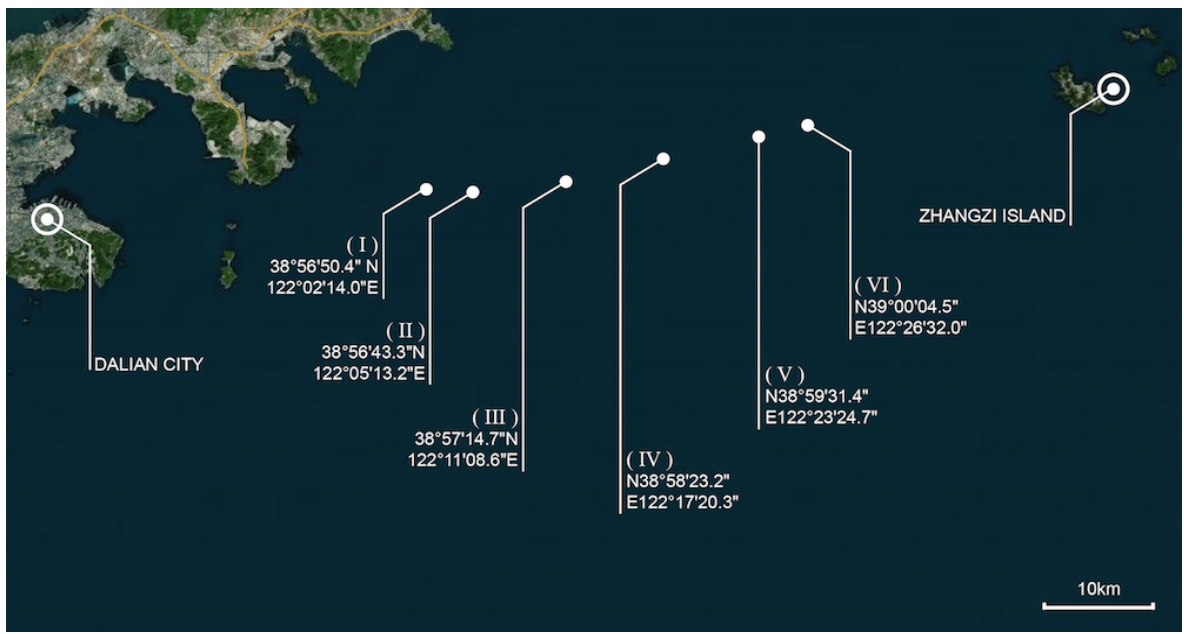

**Figure 1. Location of the seawater samples.** In order to reduce position-dependent uncertainty of seawater, we make experimental investigation in large area. The sites where we collected the samples locate at the north of the Yellow Sea, which is on the eastern coast of Liaodong peninsula, lying between Dalian city and Zhangzi island. The GPS coordinates provide specific position information of each site.

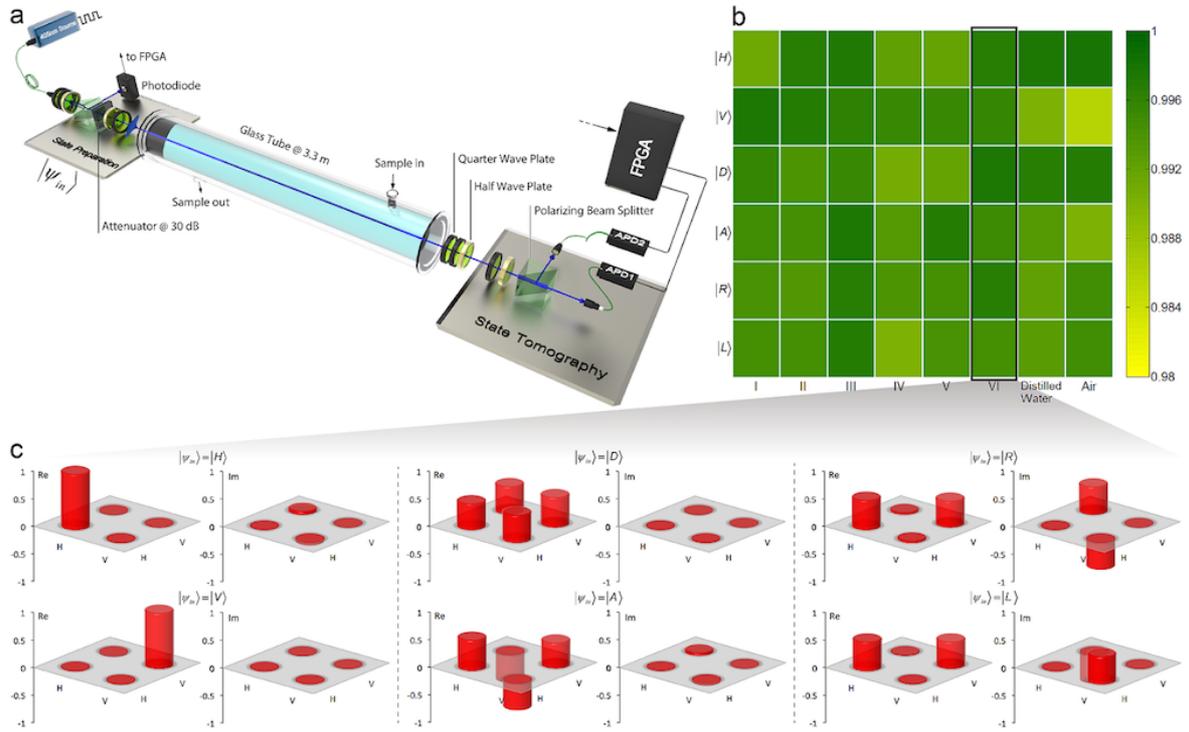

**Figure 2. Experimental quantum state transfer of single photons in free-space seawater. a**, Sketch of experimental system. The semiconductor laser can run in both continuous wave (CW) and pulsed modes. CW mode is adopted in the experiment of estimating loss of seawater samples, and pulsed mode is used to prepare single photon source with tunable attenuation. Quantum states of single photons are prepared in polarization by using a polarizing beam splitter followed by a half and a quarter wave plates, reversed order of these three elements in the output act as a state analyzer for quantum tomography. Polarization compensator consisting of two quarter- and one half-wave plates is utilized to compensate polarization rotation in fiber and other linear optical devices. **b**, 2D color chart of state ($|H\rangle$, $|V\rangle$, $|D\rangle$, $|A\rangle$, $|R\rangle$, $|L\rangle$) fidelities through different channels including six seawater samples, distilled water and air (empty tube). Maximum likelihood estimation is used for keeping density matrix physical. Note

that the start point of color bar is set at 98% to visualize the distinction better. **c**, Measured density matrix of six receiving states through seawater sample *VI*.

| Sample | I | II | III | IV | V | VI | Distilled water | Air |
|---|---|---|---|---|---|---|---|---|
| Loss (m$^{-1}$) | 0.308(10) | 0.362(10) | 0.346(9) | 0.430(32) | 0.358(8) | 0.312(18) | 0.081(9) | 0 |
| Fidelity (state) | 0.9947 | 0.9951 | 0.9968 | 0.9927 | 0.9942 | 0.9961 | 0.9939 | 0.9934 |
| Fidelity (process) | 0.9890 | 0.9926 | 0.9943 | 0.9856 | 0.9825 | 0.9912 | 0.9976 | 0.9933 |

**Table 1**. **Measured attenuation coefficients, state fidelities and process fidelities.** The results are obtained in six seawater samples (Temperature 20 °C, salinity 30‰), distilled water and air.

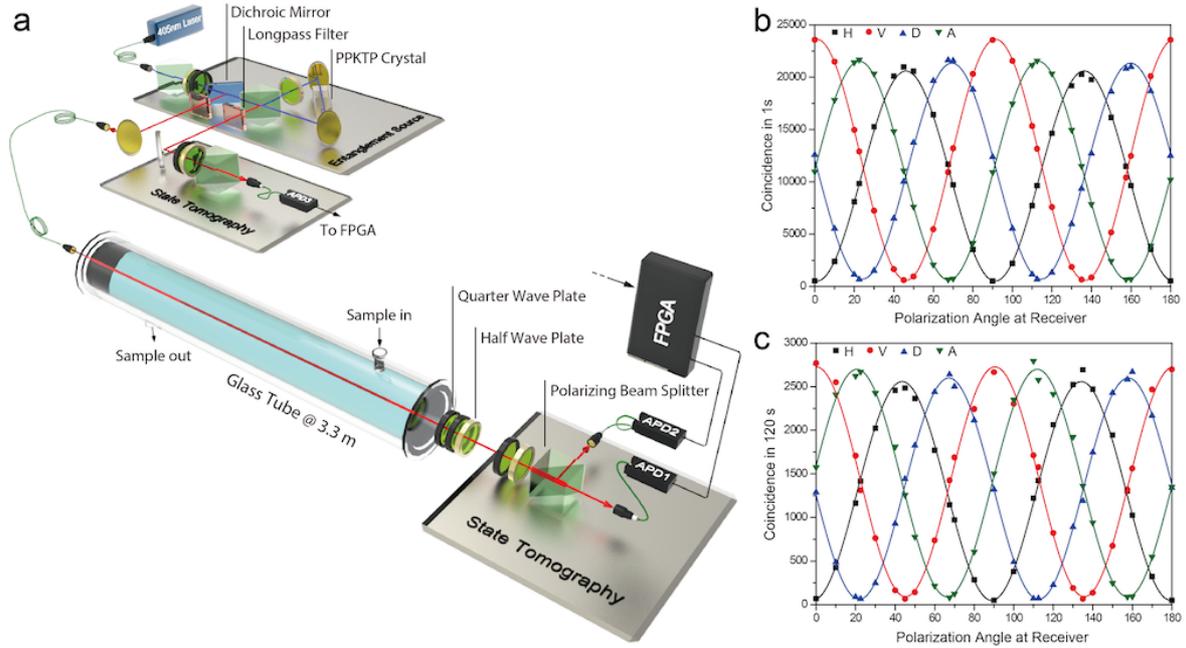

**Figure 3. Experimental entanglement distribution in free-space seawater. a**, Schematic of experimental setup. Polarization entangled photon-pair source is produced by a blue laser pumped PPKTP crystal (25-mm long) in a Sagnac ring interferometer. One photon is measured locally, the other is distributed through the channels and analyzed at the output. **b, c**, polarization correlation properties observed in air (**b**) and sample *VI* (**c**). Four curves in each chart are obtained by projecting one photon at polarization angles $|H\rangle$, $|V\rangle$, $|D\rangle$, $|A\rangle$ respectively and scan the other one.

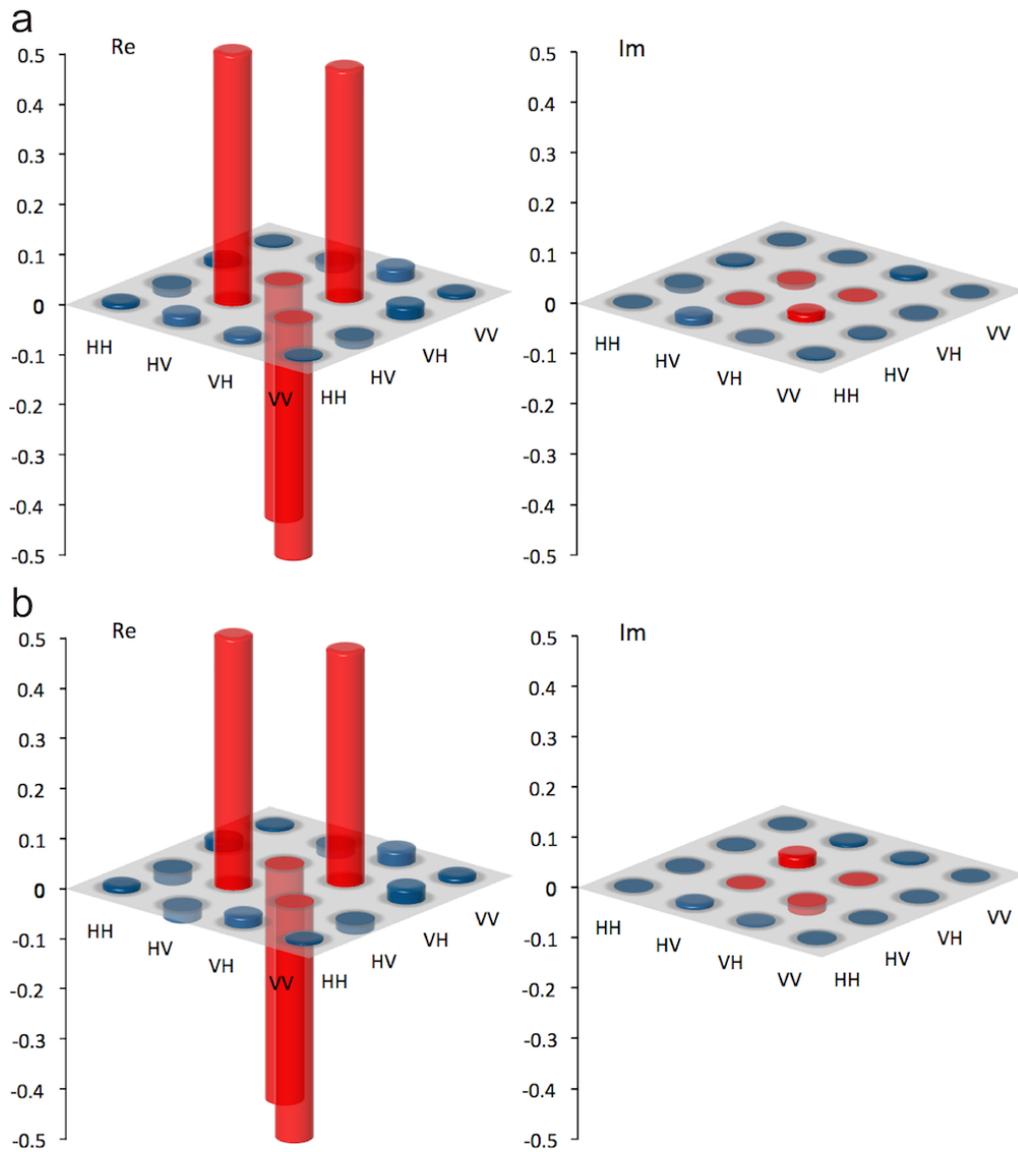

**Figure 4. Diagrammatic representation of reconstructed real (Re) and imaginary (Im) components of polarization entangled state.** Density matrix is obtained by linear state tomography under two conditions: empty tube (**a**) and sample *VI* (**b**). The states in both cases appear to be highly entangled in polarization, which demonstrate entanglement can well preserved in seawater. This complements the evidence of high process fidelities provided by Fig. 2 and Table 1.